\documentclass[3p,review]{elsarticle}

\usepackage{lineno,hyperref}
\usepackage{graphicx,amssymb,arydshln,subeqnarray,subfigure}
\usepackage{url}
\usepackage{amsbsy,amsmath,amsfonts,cases,bm,caption,color,wrapfig}
\usepackage[ruled]{algorithm2e} 
\nolinenumbers

\newcommand{\tabincell}[2]{\begin{tabular}{@{}#1@{}}#2\end{tabular}}

\journal{Digital Signal Processing}

\begin{document}

\begin{frontmatter}

\title{Neurodynamic TDOA localization with NLOS mitigation\\via maximum correntropy criterion}

\author[mymainaddress]{Wenxin~Xiong\corref{mycorrespondingauthor}}
\cortext[mycorrespondingauthor]{Corresponding author}
\ead{xiongw@informatik.uni-freiburg.de}

\author[mymainaddress]{Christian~Schindelhauer}

\author[mysecondaryaddress]{Hing~Cheung~So\fnref{fn1}}
\fntext[fn1]{EURASIP Member.}

\author[mythirdaddress]{Junli~Liang}

\author[myfourthaddress]{Zhi~Wang}

\address[mymainaddress]{Department of Computer Science, University of Freiburg, Freiburg 79110, Germany}
\address[mysecondaryaddress]{Department of Electrical Engineering, City University of Hong Kong, Hong Kong, China}
\address[mythirdaddress]{School of Electronics and Information, Northwestern Polytechnical University, Xi'an 710072, China}
\address[myfourthaddress]{State Key Laboratory of Industrial Control Technology, Zhejiang University, Hangzhou 311121, China}

\begin{abstract}
In this paper, we exploit the maximum correntropy criterion (MCC) to robustify the traditional time-difference-of-arrival (TDOA) location estimator in the presence of non-line-of-sight (NLOS) propagation conditions. For the sake of statistical efficiency, the correntropy-based robust loss is imposed on the underlying time-of-arrival composition via joint estimation of the source position and onset time, instead of the TDOA counterpart generated in the postprocessing of sensor-collected timestamps. We then employ a neurodynamic optimization approach to tackle the highly nonconvex MCC formulation. Furthermore, we examine the local stability of equilibrium for the corresponding projection-type neural network model. Simulation investigations in representative NLOS propagation scenarios demonstrate that our neurodynamic robust TDOA localization solution is capable of outperforming several existing schemes in terms of positioning accuracy.
\end{abstract}


\end{frontmatter}

\section{Introduction}
\label{Intro}
Time-difference-of-arrival (TDOA) defined as the difference in signal arrival timestamps collected at a pair of sensors removes the need for clock synchronization between the source and sensors \cite{KCHo,LLin}. It is therefore often more attractive than the time-of-arrival (TOA) based positioning in practical source localization (SL) applications \cite{FKhelifi}. A challenging issue remaining in this field, is that conventional TDOA positioning algorithms devised under the Gaussian noise assumption might nevertheless fail to work properly in the presence of outlying data. As a primary cause of outliers, non-line-of-sight (NLOS) propagation takes place when there are obstructions in the direct signal transmission path, which has in fact been widely reported in the positioning literature \cite{WXiongMCC,FYin1,PCChen,JAApo,GWang2,GWang3,ZSu,MRGholami,WXiong3,WXiong4,RMVaghefi1,JLiang2}.

\subsection{Related work}
Different types of countermeasures have been developed to mitigate the negative effects of NLOS propagation conditions in TDOA-based SL \cite{PCChen,JAApo,GWang2,GWang3,ZSu,MRGholami,WXiong3,WXiong4}.

The strategy of data selection has been mapped out in \cite{PCChen} and \cite{JAApo} to reject the outlier-prone TDOA measurements. In spite of its simplicity, false-alarms and missed-detections are generally inescapable in the implementation of data-selective schemes, not to mention the heavy computational burden. In \cite{ZSu}, the authors put forward the idea of TDOA-to-TOA model transformation (MT), in view of that the dropped source onset time can be recovered by treating it as a confined optimization variable. Just like the original study under the TOA framework in \cite{RMVaghefi1} from which \cite{ZSu} is derived, semidefinite programming is used to cope with a nonlinear least squares (LS) problem built upon the squared-range (SR) model. Benefiting from the strong resistance to outliers even when the observations are densely contaminated, worst-case based tactics have lately caught considerable attention among researchers \cite{GWang2,GWang3}. Typically, the LS formulation is robustified in the worst-case sense, whereupon convex relaxation techniques are employed to handle the resulting minimax optimization problem. Nevertheless, methods in \cite{GWang2,GWang3} hinge on additional \textit{a priori} error bounds for the realization of the robust worst-case criterion.

Another appealing way of estimator robustification is to trim the loss function of fitting errors from the statistical perspective \cite{AMZoubir}, as not explicitly introducing parameters for the error-related terms may lead to more cost-effective solutions. Since large errors can dominate the Gaussianity reliant $\ell_2$ measure and thereby resulting in wrong estimates, the authors of \cite{MRGholami} turn instead to minimizing the $\ell_1$ loss by difference-of-convex programming with a concave-convex procedure. On the other hand, the authors of \cite{WXiong3} and \cite{WXiong4} set up their outlier-resistant formulations based on the smoothed $\ell_1$-norm minimization and a generalized robust loss rooted in the locally competitive algorithm (LCA), respectively. They subsequently utilize different neurodynamic approaches to deal with the corresponding nonconvex optimization problems.

\subsection{Contribution}
Along the path of statistical robustification, this paper continues to investigate the problem of TDOA SL under NLOS propagation conditions. Our main contribution in this work is twofold:

(i) Our robust positioning solution is derived using the maximum correntropy criterion (MCC). The correntropy initially proposed for information theoretic learning \cite{JCPrincipe} is a generalized, local, and nonlinear similarity measure between two arbitrary random variables. It has lately seen tremendous growth in many fields of nonlinear and non-Gaussian signal processing \cite{WXiongMCC,JLiang2,YHe,WLiu,FDMandanas}, but has not yet been considered in the context of TDOA SL.

(ii) We apply a neurodynamic optimization approach, known as the projection-type neural network (PNN) \cite{WXiong3,HChe}, to efficiently tackle the MCC-based localization formulation. Moreover, we examine the local stability of equilibrium for the corresponding PNN model. Numerical results demonstrate that our proposed technique can achieve higher positioning accuracy than several recently developed TDOA localization methods that address the NLOS effects.

\subsection{Organization}
The remainder of the paper is organized as follows. Section \ref{Sect_MCCPF} presents our MCC-based robust TDOA SL formulation. In Section \ref{Sect_PD}, the PNN framework is established to perform the optimization, with the stability properties being later analyzed. Numerical results are included in Section \ref{Sect_NR} to evaluate the positioning performance. Finally, conclusions are drawn in Section \ref{Sect_CC}.

\section{MCC-based problem formulation}
\label{Sect_MCCPF}
\subsection{Introduction to MCC}
The correntropy between two arbitrary scalar random variables $X$ and $Y$ is defined as \cite{WLiu}
\begin{equation}{\label{Correntropy}}
V_{\sigma}(X,Y) = \mathbf{E} \left[\kappa_{\sigma} (X-Y)\right],
\end{equation}
where $\kappa_{\sigma}(x)$ is a Mercer kernel \cite{VVapnik} of size $\sigma$ and $\mathbf{E}\left[\cdot\right]$ is the expectation operator. Herein, we adopt the most commonly used Gaussian kernel function \cite{WLiu}: $\kappa_{\sigma}(x) = \exp \left(-\tfrac{x^2}{2 \sigma^2}\right)$. In the absence of the statistical knowledge about $X$ or $Y$, we may not be able to compute the expectation in (\ref{Correntropy}). For this reason, (\ref{Correntropy}) is normally replaced with the sample estimator of correntropy:
\begin{equation}{\label{SEC}}
\hat{V}_{N_{\textup{s}},\sigma} (X,Y) = \tfrac{1}{N_{\textup{s}}} \sum_{i=1}^{N_{\textup{s}}} \kappa_{\sigma} (X_i - Y_i)
\end{equation}
in practice, based on only a finite amount of available data $\left\{ X_i, Y_i \right\}_{i=1,...,N_{\textup{s}}}$.

Originally referring to maximizing the sample correntropy function (\ref{SEC}), the MCC can be alternatively attained by minimizing a decreasing function of it \cite{WLiu}. Note that the resulting correntropy-induced metric is actually strongly associated with the well-known Welsch $M$-estimator \cite{FDMandanas}.

\begin{figure}[!t]
	\centering
	\includegraphics[width=6.5in]{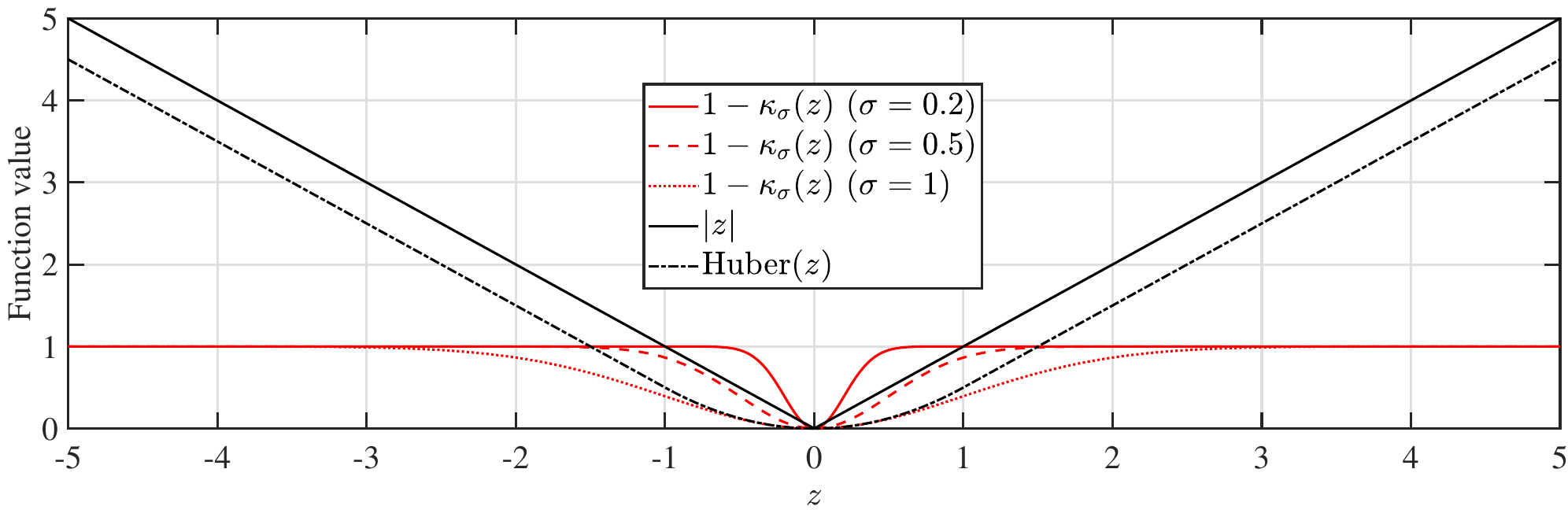}
	\caption{Comparison of robust loss functions.} 
	\label{xiong1_MCCTDOA}
\end{figure}

The MCC has an excellent and very appealing benefit that all the properties of correntropy (i.e., the robustness of correntropy-based loss function in our case) are \textit{smoothly} controlled by the kernel size $\sigma$ \cite{WLiu}. As an illustration, we provide in Fig. \ref{xiong1_MCCTDOA} a comparison of different robust loss functions in the one-dimensional space, namely, $1-\kappa_{\sigma} (z)$, $|z|$, and $\textup{Huber}(z)=\left\{ \begin{aligned} &(1/2)z^2,~|z| \leq 1\\&|z| - 1/2,~|z| > 1 \end{aligned} \right.$ with unit Huber radius. It is seen that fitting errors of abnormally large magnitude can be effectively mitigated via the nonconvex correntropy measure by selecting a proper $\sigma$. On the other hand, convex loss functions like the $\ell_1$ and Huber's can only in part limit the influence of outliers. This is somewhat analogous to the dichotomy between monotone and redescending estimators within the $M$-estimator framework \cite{CCLee}.

\subsection{Problem formulation}
Our localization scenario consists of $L \geq H+1$ coordinated sensors and a single source to be located in the $H$-dimensional space ($H=2$ or $3$). The known position of the $i$th sensor and unknown source location are denoted by $\bm{x}_i \in \mathbb{R}^{H}$ (for $i = 1,...,L$) and $\bm{x} \in \mathbb{R}^{H}$, respectively. In the passive mode, a radio or sound signal is emitted from the source at the onset time $t_0 > 0$ and received by the $i$th sensor (for $i = 1,...,L$) at time $t_i > t_0$ afterwards (namely, $t_i$ corresponds to the received signal timestamp available from the $i$th sensor).

The TOA between the $i$th sensor and source is modeled as
\begin{equation}{\label{measmdl}}
t_i - t_0 = \tfrac{1}{c}\left({\|\bm{x} - \bm{x}_i\|}_2 + n_i + q_i \right),~i = 1,...,L,
\end{equation}
where $c$ is the signal propagation velocity, ${\|\cdot\|}_2$ represents the $\ell_2$-norm of a vector, $n_i$ denotes the measurement noise due to thermal disturbance at the $i$th sensor and is an independent zero-mean Gaussian process with variance $\sigma_{i}^2$, and $q_i$ is either 0 or a positive bias error, depending on whether the corresponding source-sensor path is under line-of-sight (LOS) or NLOS propagation. For simulation purposes, the possibly biased $q_i$ is typically modeled as uniform distribution \cite{WXiongMCC,JAApo,GWang2,GWang3,ZSu,WXiong3,WXiong4,WXiong,HChen2}, shifted Gaussian distribution \cite{FYin1}, or exponential distribution \cite{PCChen,RMVaghefi1} in the literature. Nonetheless, to be more realistic, no prior knowledge is assumed about the statistics of noise/errors or error status in the problem-solving stage (i.e., such information is unknown to our positioning algorithm).

However, due to the lack of clock synchronization between the source and any sensor in our TDOA setting, $t_0$ is unknown and therefore (\ref{measmdl}) is unavailable. The following TDOAs are often calculated as an alternative:
\begin{align}{\label{TDOA}}
t_{i,j} &= \tfrac{1}{c}({\| \bm{x} - \bm{x}_i \|}_2 - {\| \bm{x} - \bm{x}_j \|}_2 + n_{i,j} + q_{i,j})\nonumber\\
&= t_i - t_j,~~i \not= j,~i,j = 1,...,L,
\end{align}
where $n_{i,j} = n_i - n_j$ and $q_{i,j} = q_i - q_j$ (both for $i \not= j, i,j = 1,...,L$) are the noise and error terms in the corresponding TDOA-based range-difference (RD) measurement.

The task of robust TDOA SL is to reliably estimate $\bm{x}$ given $\{ \bm{x}_i \}$, $c$, and possibly erroneous $\{ t_i \}_{i=1,...,L}$ (or $\{ t_{i,j} \}$ in (\ref{TDOA}), generated in the postprocessing of sensor-collected timestamps).

A challenging issue is that the traditional $\ell_2$ (LS) based TDOA location estimator in general performs badly in the presence of outliers \cite{GWang2}. Straightforwardly, the concept of robust statistics can be borrowed to overcome the drawback. Designating the first sensor as the reference, the authors of \cite{MRGholami,WXiong4} focus on $L-1$ nonredundant RDs $\{ ct_{i,1} \}$ and suggest handling:
\begin{equation}\label{robust_form}
	\min_{\bm{x}} \sum_{i=2}^{L} \psi (e_{i,1}),
\end{equation}
where the dummy variable for the fitting error, $e_{i,1}$, satisfies $e_{i,1} = c t_{i,1} - {\| \bm{x} - \bm{x}_i \|}_2 + {\| \bm{x} - \bm{x}_1 \|}_2$ (for $i = 2,...,L$), and $\psi(\cdot)$ is a cost function robust to outliers.

Similarly, one may follow the way in (\ref{robust_form}) and let $\psi(\cdot)$ be a decreasing function of the sample correntropy to achieve statistical robustification in the MCC sense. However, this kind of robustification potentially suffers from the issue that imposing error measures on (\ref{TDOA}) possesses lower statistical efficiency than directly applying them to (\ref{measmdl}) \cite{GWang3,ZSu,WXiong3}. Taking it into consideration when formulating our MCC problem, we treat $t_0$ as an additional optimization variable \cite{ZSu} and robustify the TDOA location estimator based on the underlying TOA composition:
\begin{align}{\label{robust_form2}}
\min_{t_0,\bm{x}} \sum_{i=1}^{L} [-\kappa_{\sigma} (e_{i})],
\end{align}
where $e_{i} = c(t_i - t_0) - {\|\bm{x} - \bm{x}_i\|}_2$ (for $i=1,...,L$).

\section{PNN design}
\label{Sect_PD}
Prior to handling (\ref{robust_form2}), we point out that several reasonable constraints should be taken into account for ensuring the boundedness of the newly introduced estimation variable $t_0$ \cite{ZSu,WXiong3}. This brings us the following constrained minimization substitute for (\ref{robust_form2}):
\begin{subequations}{\label{robust_form2re}}
\begin{align}
\min_{t_{0},\bm{x},\bm{d}}&~\sum_{i=1}^{L} \Big\{ -\kappa_{\sigma} [ c(t_i - t_{0}) - d_i ] \Big\} \nonumber\\
\textup{s.t.}~~&0 \leq t_0 \leq t_i,~~i = 1,...,L,\label{robust_form2rea}\\
&d_i^2 = {\|\bm{x} - \bm{x}_i\|}_2^2,~d_i \geq 0,~~i = 1,...,L,\label{robust_form2reb}\\
&(t_i - t_0)c + (t_j - t_0)c \geq {\| \bm{x}_i - \bm{x}_j \|}_2,~~i \not= j,~i,j = 1,...,L,\label{robust_form2rec}\\
&(t_i - t_0) c \geq d_i,~~i = 1,...,L,\label{robust_form2red}
\end{align}
\end{subequations}
where (\ref{robust_form2rea}), (\ref{robust_form2rec}), and (\ref{robust_form2red}) are the temporal constraints, geometrical constraints by the triangle inequality, and constraints based on the general consensus that $q_i$ is normally much greater than $|n_i|$, respectively \cite{ZSu,WXiong3}. It is noteworthy that (\ref{robust_form2rec}) and (\ref{robust_form2red}) may not be satisfied in certain cases where $n_i$ is negative and $q_i$ is of small magnitude. This might result in an infeasible program if one resorts to convex relaxation techniques \cite{ZSu}. Instead, we take a neurodynamic optimization approach to handle the constrained minimization formulation in (\ref{robust_form2re}).

The use of analog neural networks for optimization has a long history, dating back more than three decades to the pioneering work of Hopfield and Tank \cite{DTank}. Consider the following paradigm of a constrained minimization problem without any convexity assumptions:
\begin{align}{\label{Gen_Cons_P}}
\min_{\bm{y} \in \mathbb{R}^N}~f(\bm{y}),\quad\textup{s.t.}~~\bm{g}(\bm{y}) \leqq \bm{0}_K,~~\bm{h}(\bm{y}) = \bm{0}_M,
\end{align}
where $f:\mathbb{R}^N \rightarrow \mathbb{R}$, $\bm{g}(\bm{y}) = \left[ g_1 (\bm{y}),...,g_K (\bm{y}) \right]^T \in \mathbb{R}^K$ and $\bm{h}(\bm{y}) = \left[ h_1 (\bm{y}),...,h_M (\bm{y}) \right]^T \in \mathbb{R}^M$ are the $K$- and $M$-dimensional vector-valued functions of $N$ variables, respectively, $f(\bm{y})$ and $h_i (\bm{y})$ (for $i = 1,...,M$) are all differentiable, $\bm{0}_K \in \mathbb{R}^K$ denotes the $K \times 1$ all-zero vector, the vector inequality $\bm{a} \leqq \bm{b}$ means $[\bm{a}]_{i} \leq [\bm{b}]_{i}$ for all choices of $i$, and $[\cdot]_{i} \in \mathbb{R}$ represents the $i$th element of a vector.

In cases where $f(\bm{y})$ or $g_i (\bm{y})$ is nonconvex, or $h_i (\bm{y})$ is not an affine expression, (\ref{Gen_Cons_P}) does not constitute a disciplined convex programming problem and finding the global minimizer can be difficult. Comparatively speaking, it would be easier to employ certain locally stable neurodynamic optimization methods \cite{WXiong3,DTank,ANazemi,SZhang,ZHan,HChe}, whose equilibrium state is reached at a Karush-Kuhn-Tucker (KKT) point (viz., a point satisfying the first-order necessary conditions of optimality) of (\ref{Gen_Cons_P}). In order to directly take the inequality constraints into account and avoid introducing dummy variables, we turn to a PNN solution \cite{WXiong3,HChe}, derived based on the projection theorem and a redefined augmented Lagrangian of (\ref{Gen_Cons_P}):
\begin{equation}
	\mathcal{L}_{\rho}(\bm{y},\bm{\upsilon}) = f(\bm{y}) + \bm{\beta}^T \bm{g}(\bm{y}) + \bm{\gamma}^T \bm{h}(\bm{y}) + \frac{\rho}{2} \left\{ \sum_{i = 1}^{K} \left[ \beta_i g_i (\bm{y}) \right]^2 + \sum_{i = 1}^{M} \left[ \gamma_i h_i (\bm{y}) \right]^2 \right\},
\end{equation}
where $\bm{\upsilon} = \left[ \bm{\beta}^T, \bm{\gamma}^T \right]^T \in \mathbb{R}^{K+M}$, $\bm{\beta} = \left[ {\beta}_1,...,{\beta}_K \right]^T \in \mathbb{R}^K$ and $\bm{\gamma} = \left[ {\gamma}_1,...,{\gamma}_M \right]^T \in \mathbb{R}^M$ are vectors containing Lagrange multipliers for the inequality constraints and equality constraints in (\ref{Gen_Cons_P}), respectively, and $\rho > 0$ is the augmented Lagrangian parameter. The time-domain transient behaviors of the PNN are governed by \cite{WXiong3,HChe}
\begin{align}{\label{dynamicsGen_Cons_P}}
\frac{d\bm{y}}{dt} = -\bm{\nabla}_{\bm{y}} \mathcal{L}_{\rho}(\bm{y},\bm{\upsilon}),~~\frac{d{\beta_i}}{dt} = -\beta_i + \left[ \beta_i + g_i (\bm{y}) \right]^{+},~i = 1,...,K,~~\frac{d{\bm{\gamma}}}{dt} = \bm{h}(\bm{y}),
\end{align}
where $\bm{\nabla}_{\bm{y}} (\cdot) = \left[ \tfrac{\partial(\cdot)}{\partial t_{0}}, \left( \tfrac{\partial (\cdot)}{\partial \bm{x}} \right)^T, \left( \tfrac{\partial (\cdot)}{\partial \bm{d}} \right)^T \right]^T \in \mathbb{R}^N$ calculates the gradient of a function at $\bm{y}$, and $\left[ \cdot \right]^{+} = \max (\cdot, 0)$ actually defines a nonlinear projection that acts like the unit ramp function.

\begin{figure}[!t]
	\centering
	\includegraphics[width=6.5in]{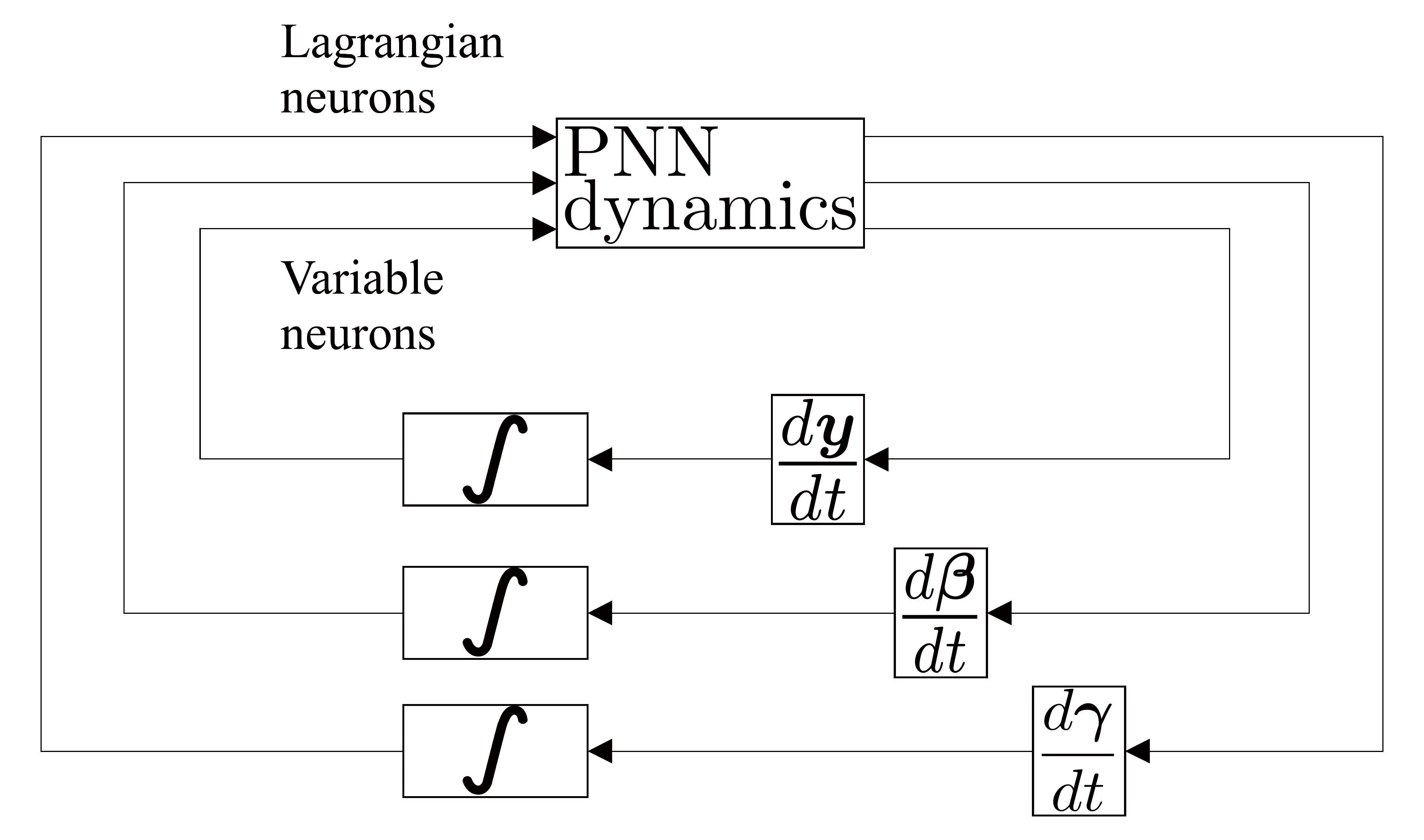}
	\caption{Realization of PNN.} 
	\label{PNN_hard}
\end{figure}

Fig. \ref{PNN_hard} gives an illustration of how the PNN is realized. Similar to the well-known Lagrange programming neural network \cite{WXiong4,SZhang,ZHan}, the PNN is composed of two types of simultaneously contributing analog computational elements, namely, the variable neurons and Lagrangian neurons, which hold the optimization variables in $\bm{y}$ and Lagrange multipliers in $\bm{\upsilon}$, respectively. Described by the dynamical equations in (\ref{dynamicsGen_Cons_P}), the physical meaning of neuronal activities is assigned to $\bm{y}$ and $\bm{\upsilon}$, in a manner that $\frac{d\bm{y}}{dt} = -\bm{\nabla}_{\bm{y}} \mathcal{L}_{\rho}(\bm{y},\bm{\upsilon})$ renders $\mathcal{L}_{{\rho}}(\bm{y},\bm{\upsilon})$ decreasing, and $\frac{d{\beta_i}}{dt} = -\beta_i + \left[ \beta_i + g_i (\bm{y}) \right]^{+}$ as well as $\frac{d{\bm{\gamma}}}{dt} = \bm{h}(\bm{y})$ aim at guiding the trajectory into the feasible region of (\ref{Gen_Cons_P}). Once the equilibrium is achieved, the solution to (\ref{Gen_Cons_P}) is obtained from the neurons associated with $\bm{y}$.

Back to our MCC-based robust TDOA SL problem, the constrained minimization formulation in (\ref{robust_form2re}) coincides with (\ref{Gen_Cons_P}) as long as we have the following redefinition:

\begin{align}
	&N = H+L+1,\\
	&\bm{y} = \left[ t_0, \bm{x}^T,\bm{d}^T \right]^T,\\
	&f(\bm{y}) = \sum_{i=1}^{L} \left\{ -\kappa_{\sigma} [ c(t_i - t_{0}) - d_i ] \right\},\\
	&K = \tfrac{L^2+5L}{2} + 1,\\
	&g_1 (\bm{y}) = - t_0,\\
	&g_{i+1} (\bm{y}) = t_0 - t_i,~~i = 1,...,L,\\
	&g_{i+L+1} (\bm{y}) = -d_i,~~i = 1,...,L,\\
	&g_{i+2L+1} (\bm{y}) = d_i - (t_i - t_0)c,~~i = 1,...,L,\\
	&g_{[(2L-i)(i-1)/2]+1+3L+j-i} (\bm{y}) = g_{i,j} (\bm{y}) = {\| \bm{x}_i - \bm{x}_j \|}_2 - (t_i - t_0)c - (t_j - t_0)c,\nonumber\\
	&~~i = 1,...,L-1,~j = i+1, i+2,...,L,\\
	&M = L,\\
	&h_i (\bm{y}) = d_i^2 - {\| \bm{x} - \bm{x}_i \|}_2^2,~~i = 1,...,L.
\end{align}

For ease of exposition, the neurodynamic TDOA-based localization scheme presented in this paper is succinctly termed MCC-PNN. Just as its name implies, MCC-PNN leverages the robust correntropy-induced loss to reduce the adverse effects of unreliable sensor-collected timestamps and, in particular, a PNN is applied to deal with the related nonconvex constrained optimization problem.

Several relatively more complicated gradient calculations in (\ref{dynamicsGen_Cons_P}) are detailed as follows:

\begin{align}{\label{detailed_dynamics}}
	&\frac{\partial \mathcal{L}_{\rho} (\bm{y},\bm{\upsilon})}{\partial t_0} = \sum_{i=1}^{L} \frac{c \left[ d_i - (t_i - t_0) c \right]}{\sigma^2} \exp \left\{ -\frac{{\left[(t_i - t_0) c - d_i \right]}^2}{2 \sigma^2} \right\} - \beta_1 + \sum_{i=1}^{L} \beta_{i+1} + c \sum_{i=1}^{L} \beta_{i+2L+1} \nonumber\\
	&~~+ 2c \sum_{i=1}^{L-1} \sum_{j=i+1}^{L} \beta_{[(2L-i)(i-1)/2]+1+3L+j-i} + \rho \Bigg\{ -\beta_1^2 g_1 (\bm{y}) + \sum_{i=1}^{L} \beta_{i+1}^2 g_{i+1} (\bm{y}) + c \sum_{i=1}^{L} \beta_{i+2L+1}^2 g_{i+2L+1} (\bm{y}) \nonumber\\
	&~~+ 2c \sum_{i=1}^{L-1} \sum_{j=i+1}^{L} \beta_{[(2L-i)(i-1)/2]+1+3L+j-i}^2 g_{[(2L-i)(i-1)/2]+1+3L+j-i} (\bm{y}) \Bigg\}, \nonumber\\
	&\frac{\partial \mathcal{L}_{\rho} (\bm{y},\bm{\upsilon})}{\partial \bm{x}} = 2 \sum_{i=1}^{L} \left( \bm{x}_i - \bm{x} \right) \left[ \rho \gamma_{i}^2 h_i (\bm{y}) + \gamma_{i} \right], \nonumber\\
	&\frac{\partial \mathcal{L}_{\rho} (\bm{y},\bm{\upsilon})}{\partial d_i} = \frac{ \left[ d_i - (t_i - t_0) c \right]}{\sigma^2} \exp \left\{ -\frac{{\left[(t_i - t_0) c - d_i \right]}^2}{2 \sigma^2} \right\} - \beta_{i+L+1} +\beta_{i+2L+1} - 2 \gamma_{i} g_{i+L+1} (\bm{y}) \nonumber\\
	&~~+ \rho \left[ -\beta_{i+L+1}^2 g_{i+L+1} (\bm{y}) + \beta_{i+2L+1}^2 g_{i+2L+1} (\bm{y}) + 2 \gamma_{i}^2 d_i h_i (\bm{y}) \right],~~i = 1,...,L.
\end{align}

According to the previous research in \cite{WXiong3,HChe}, PNN governed by (\ref{dynamicsGen_Cons_P}) is assured locally stable, namely, it will reach equilibrium at a KKT point $\left(\bm{y}^{*}, \bm{\upsilon}^{*}\right)$ for (\ref{Gen_Cons_P}) under very mild conditions, where $\bm{y}^{*}$ is a strict local minimum of (\ref{Gen_Cons_P}). With a large enough\footnote{A typical choice of value for the augmented Lagrangian parameter can be, say, $\rho = 5$ \cite{WXiong3} or $\rho = 10$ \cite{WXiong}. Particularly, the former is known to be able to accelerate the transition but without leading to divergence \cite{SZhang}.} $\rho$, the sufficient conditions for its local stability can be summarized as \cite{WXiong3}: (i) the Hessian of the restricted Lagrangian at a KKT point is positive definite on the critical cone, and (ii) assume that $\bm{y}^{*}$ is a regular point, the gradients of equality constraints and active inequality constraints w.r.t. $\bm{y}$ are linearly independent.

Let us focus first on the verification of the second condition. Under the assumption that the source location in general does not overlap sensor positions (otherwise $d_i^{*} = 0$ and there would be no need for localization), it is deemed none of the inequality constraints in (\ref{Gen_Cons_P}) are active. Therefore, we may only calculate the gradient of $\bm{h}(\bm{y})$ w.r.t. $\bm{y}$ at $\left(\bm{y}^{*}, \bm{\upsilon}^{*}\right)$ for verifying the second condition\footnote{For notational convenience, we assume that the asterisk in the superscript of a vector applies to each element of that vector by default.}, i.e.
\begin{align}{\label{Gradh_MCCTDOA2}}
\bm{\nabla}_{\bm{y}} \bm{h}(\bm{y}^{*}) = \begin{bmatrix}
\begin{array}{c | c | c}
\bm{0}_L  &  2 \bm{X}^T - 2 \bm{1}_L {\bm{x}^{*}}^T  &  2\mathrm{diag}(\bm{d}^*)
\end{array}
\end{bmatrix},
\end{align}
where $\bm{X} = \left[ \bm{x}_1,..., \bm{x}_L \right] \in \mathbb{R}^{H \times L}$ and $\mathrm{diag}(\cdot)$ is a diagonal matrix with the corresponding vector being the main diagonal. We see that the row vectors of the matrix in (\ref{Gradh_MCCTDOA2}) are linearly independent, again on the premise of $d_i^{*} \neq 0$ (for $i = 1,...,L$). Combining the inactivity of inequality constraints and linear independence just verified further deduces an empty critical cone \cite{WXiong3,HChe}. Consequently, two conditions are both met and the local stability of MCC-PNN is confirmed.

\section{Numerical results}
\label{Sect_NR}
In this section, we carry out numerical investigations to substantiate the efficacy of MCC-PNN.

In our simulations, the variance of the zero-mean white Gaussian noise $n_i$, $\sigma_i^2$, is assumed to be of a constant value $0.1~\textup{m}^2$ for all choices of $i$. The possible error $q_i$ is randomly drawn from a uniform distribution $\mathcal{U}(0, b_i)$ with parameter $b_i \geq 0$, namely, the $i$th source-sensor path is under LOS propagation if $b_i = 0$ and NLOS otherwise. The number of source-sensor paths under NLOS propagation conditions is denoted by $L_{\textup{NLOS}}$. For simplicity, we have $\{ b_i = b \}_{i=1,...,L_{\textup{NLOS}}}$ and $\{ b_i = 0 \}_{i=L_{\textup{NLOS}}+1,...,L}$ for source-sensor paths in the NLOS and LOS scenarios, respectively. By conducting a total of 500 Monte Carlo (MC) runs, the root mean square error (RMSE), calculated as $\textup{RMSE} = \sqrt{\frac{1}{500}\sum_{i=1}^{500}{{\left\|\hat{\bm{x}}^{\{i\}} - \bm{x}^{\{i\}}\right\|}_2^2}}$, is employed to assess the positioning performance, where $\hat{\bm{x}}^{\{i\}}$ represents the estimate of the source position $\bm{x}^{\{i\}}$ in the $i$th MC trial.

It is worth pointing out that the PNN is an analog neural computational technique, initially intended for physical realization by designated hardware (e.g., application-specific integrated circuits \cite{ANazemi}). Nonetheless, the dynamical system in (\ref{dynamicsGen_Cons_P}) can also be implemented in a numerical fashion, either with the use of the MATLAB ordinary differential equation (ODE) solver \cite{WXiong4,ZHan,LFShampine}, or by following the procedure below for discrete realization \cite{JLiang2}:
\begin{equation}{\label{DiscReal}}
	\left\{ \begin{aligned} &\bm{y}^{(k+1)} = \bm{y}^{(k)} + \tau \left. \frac{d\bm{y}}{dt} \right|_{\bm{y} = \bm{y}^{(k)}}\\&\bm{\beta}^{(k+1)} = \bm{\beta}^{(k)} + \tau \left. \frac{d\bm{\beta}}{dt} \right|_{\bm{\beta} = \bm{\beta}^{(k)}}\\&\bm{\gamma}^{(k+1)} = \bm{\gamma}^{(k)} + \tau \left. \frac{d\bm{\gamma}}{dt} \right|_{\bm{\gamma} = \bm{\gamma}^{(k)}} \end{aligned} \right.,
\end{equation}
where the iterations are indexed by $(\cdot)^{(k)}$ and $\tau>0$ is the step size. Towards a trade-off between the convergence speed, numerical stability, and precision, we fix the step size as $\tau = 0.0001$ in this paper.

\begin{figure*}[!t]
	\centering
	\subfigure[]{\includegraphics[width=2.7in]{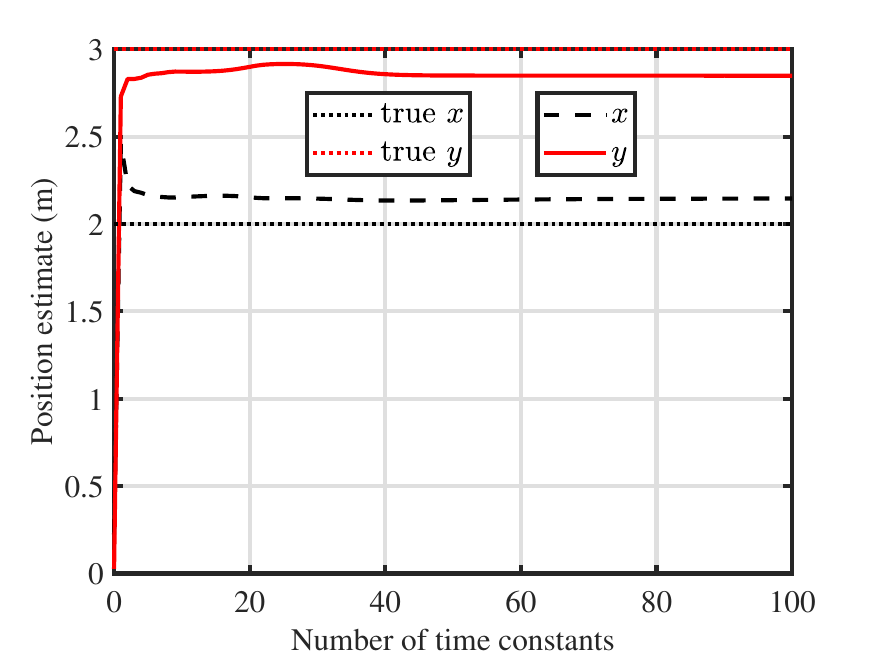}}
	\subfigure[]{\includegraphics[width=2.7in]{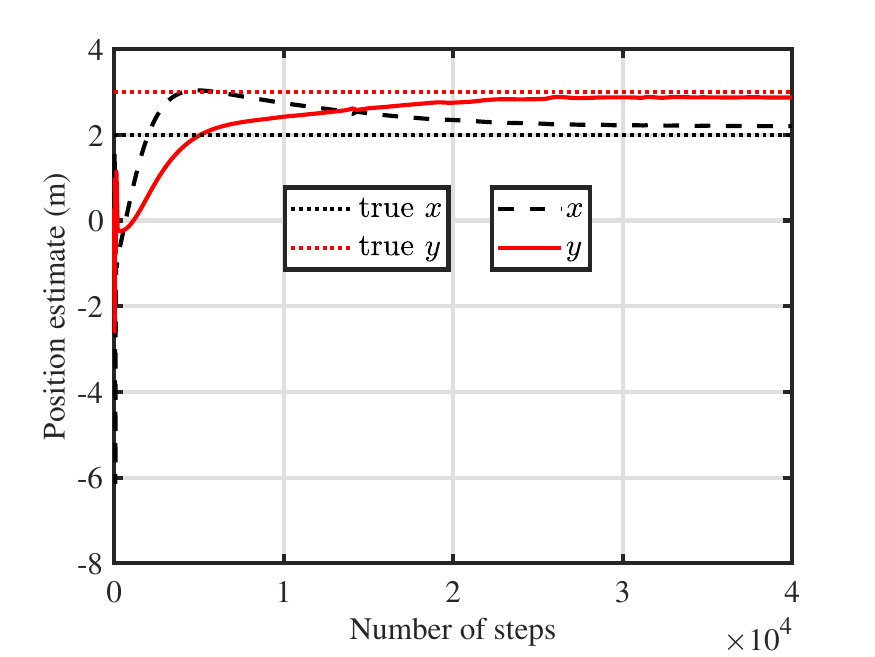}}
	\subfigure[]{\includegraphics[width=2.7in]{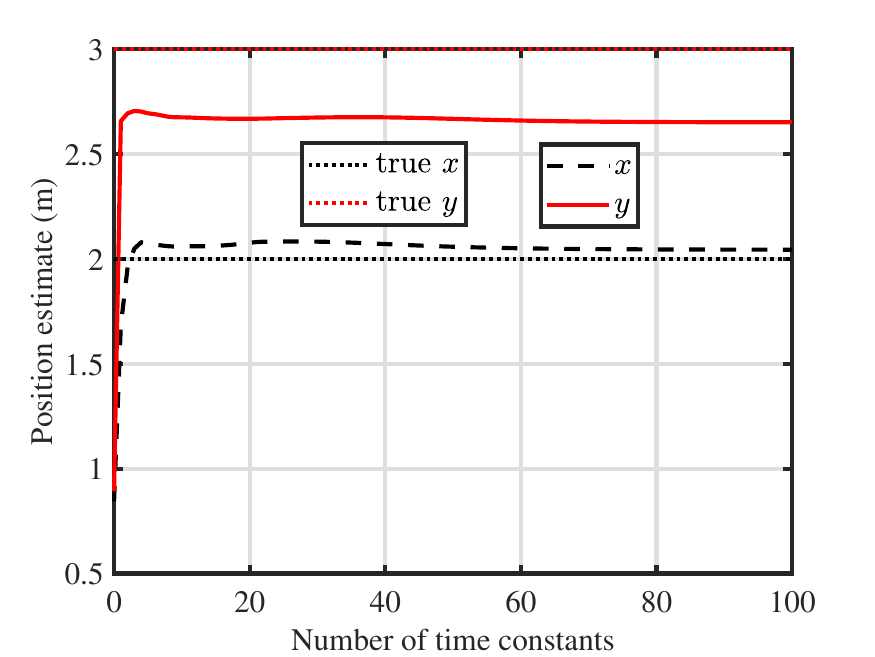}}
	\subfigure[]{\includegraphics[width=2.7in]{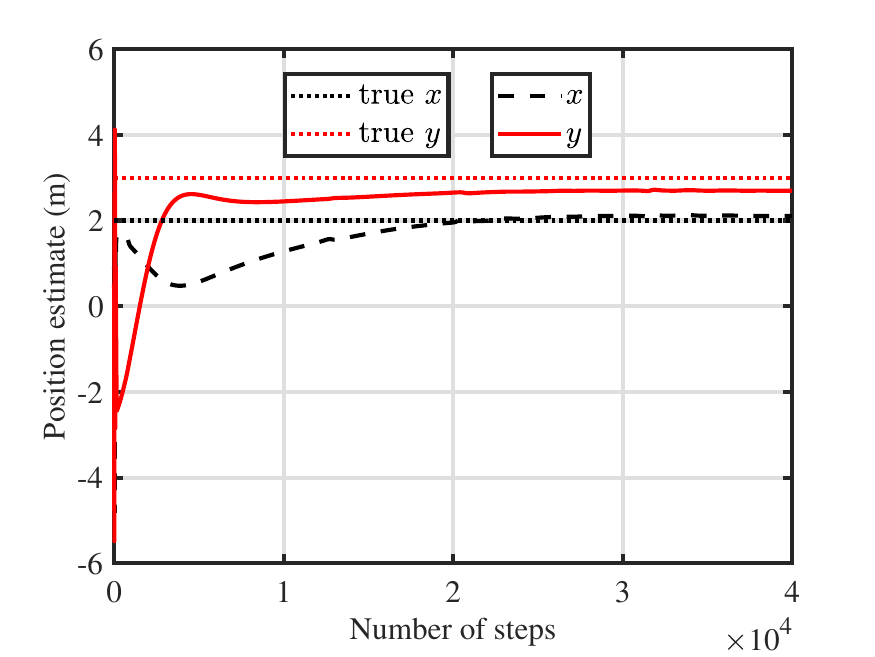}}
	\subfigure[]{\includegraphics[width=2.7in]{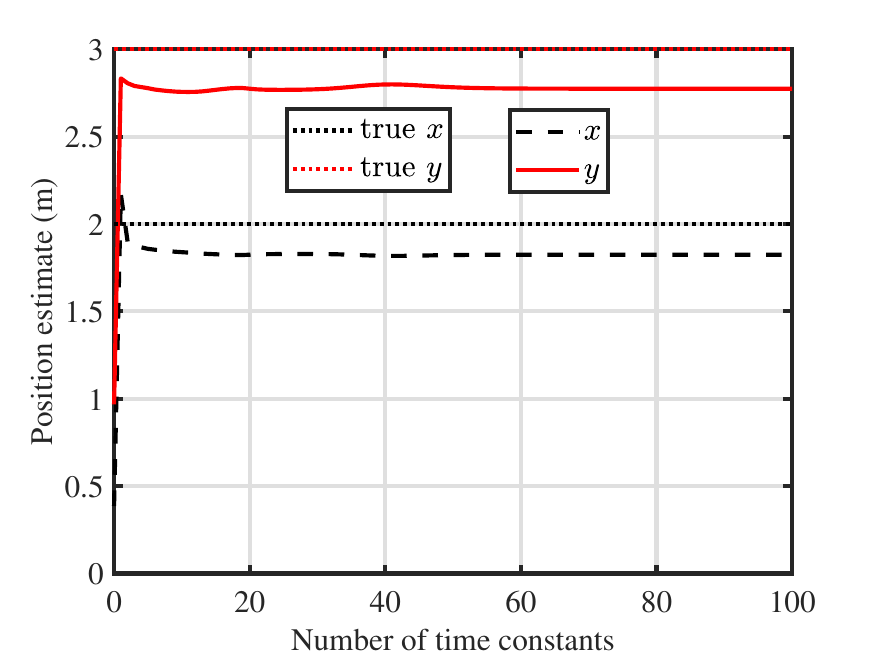}}
	\subfigure[]{\includegraphics[width=2.7in]{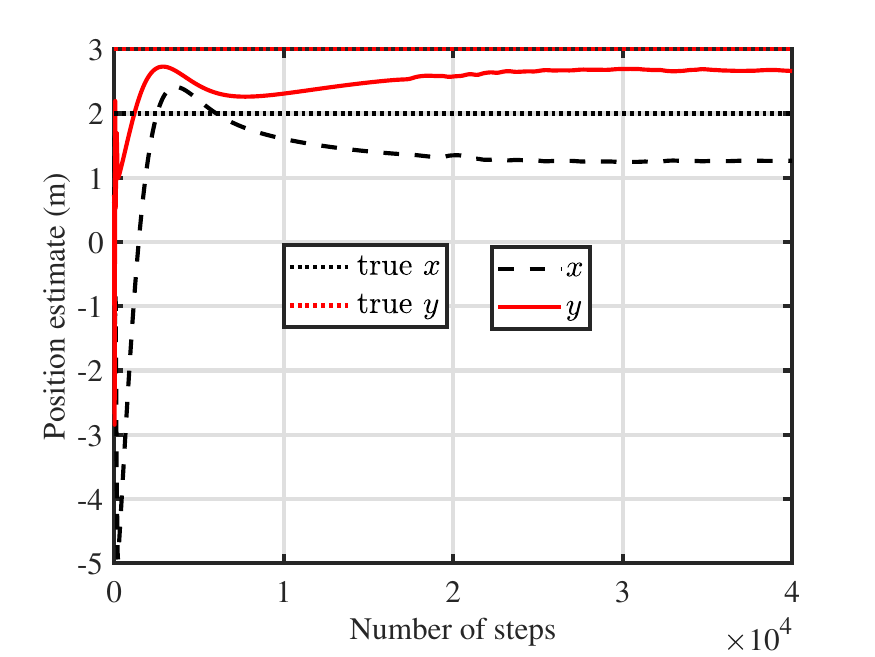}}
	\centering
	\caption{Dynamic behaviors of position estimate versus number of time constants/steps for deterministic deployment in mild, moderate, and severe NLOS environments. (a) ODE implementation with $L_{\textup{NLOS}} = 2$. (b) Discrete realization with $L_{\textup{NLOS}} = 2$. (c) ODE implementation with $L_{\textup{NLOS}} = 4$. (d) Discrete realization with $L_{\textup{NLOS}} = 4$. (e) ODE implementation with $L_{\textup{NLOS}} = 6$. (f) Discrete realization with $L_{\textup{NLOS}} = 6$.}
	\label{Figs_DBehaviors}
\end{figure*}

For illustrative purposes, let us consider first a deterministic deployment of source and sensors with $H = 2$, $L = 8$, $\bm{x} = [2, 3]^T$ m, and $\bm{X} = \begin{bmatrix} -10 & 0 & 10 & 10 & 10 & 0 & -10 & -10 \\ 10 & 10 & 10 & 0 & -10 & -10 & -10 & 0 \end{bmatrix}$ m. Fig. \ref{Figs_DBehaviors} shows the dynamic behaviors of source location estimate for both ODE implementation and discrete realization of MCC-PNN in a single MC run with $b = 5$. Here, the mild ($L_{\textup{NLOS}} = 2$), moderate ($L_{\textup{NLOS}} = 4$), and severe ($L_{\textup{NLOS}} = 6$) NLOS environments are covered. The kernel width $\sigma$ is set to $0.8$ for the moment. Values held in the neurons are randomly initialized. We see in Fig. \ref{Figs_DBehaviors} that an equilibrium point is reached within either several tens of ODE time constants, or several tens of thousands of iterations by (\ref{DiscReal}). In the rest of the section, we will implement MCC-PNN by means of (\ref{DiscReal}), which provides a more general solution.

In another sense, the procedure in (\ref{DiscReal}) enables us to quantify the computational complexity of MCC-PNN. Our quantification is based on the assumption that the update of values held in neurons dominates the computational cost of each iteration, and the fact that the evaluation operation of a degree-$n$ polynomial with fixed-size coefficients using Horner's method \cite{EHildebrand} has a complexity of $\mathcal{O}(n)$. As a consequence, the total complexity of MCC-PNN is $\mathcal{O}\left(N_{\text{PNN}}L^2\right)$, where $N_{\text{PNN}}$ denotes the number of iterations taken in discretely realizing the PNN.

As we have mentioned in Section \ref{Sect_MCCPF}, a crucial issue in regard to the MCC-based robustification is how should one prudently select the performance-decisive kernel size $\sigma$. Roughly speaking, a relatively small $\sigma$ may result in higher estimation accuracy, and the MCC performs outstandingly when $\sigma$ is in the range of $[0.2, 2]$ (see the early investigations in \cite{WLiu}). Apart from the fixed-value scheme, there exist also adaptive updating rules with the kernel size being adjusted at each iteration, which possess better flexibility \cite{WLiu}. We adopt herein the celebrated Silverman's heuristic \cite{BWSilverman}:
\begin{equation}
	\sigma^{(k+1)} = 1.06 \times \min \{ \sigma_{E}^{(k+1)}, R^{(k+1)}/1.34 \} \times L^{-0.2},
\end{equation}
where $\sigma_{E}$ and $R$ represent the standard deviation and interquartile range of $\{ e_i \}_{i=1,...,L}$, respectively, and the kernel width is initialized as $\sigma^{(0)} = 10000$ \cite{YHe}.

\begin{table*}[!t]
	\renewcommand{\arraystretch}{1}
	\caption{Summary of TDOA SL techniques being considered for comparison.}
	\label{table_MCCTDOA}
	\centering
	\begin{tabular}{|c|c|c|}
		\hline
		\bfseries Method & \bfseries Input & \bfseries Complexity\\
		\hline
		MCC-PNN & \tabincell{c}{Sensor positions\\Received signal timestamps\\Signal propagation speed} & $\mathcal{O}\left(N_{\text{PNN}}L^2\right)$\\
		\hline
		SDP-TOA \cite{ZSu} & \tabincell{c}{Sensor positions\\Received signal timestamps\\Signal propagation speed} & $\mathcal{O}\left(L^{4}\right)$\\
		\hline
		SDP-Robust-R1 \cite{GWang3} & \tabincell{c}{Sensor positions\\TDOA-based RD measurements\\Upper bounds on NLOS errors} & $\mathcal{O}\left(L^{6.5}\right)$\\
		\hline
		SDP-Robust-R2 \cite{GWang3} & \tabincell{c}{Sensor positions\\TDOA-based RD measurements\\Upper bounds on NLOS errors} & $\mathcal{O}\left(L^{6.5}\right)$\\
		\hline
	\end{tabular}
\end{table*}

We now proceed to compare the performance of MCC-PNN with several recently proposed TDOA-based SL schemes in two random deployment scenarios. The first configuration continues to use the aforementioned $L=8$ sensors with fixed positions, but the single source is randomly and uniformly distributed in the origin-centered 20 m $\times$ 20 m square region (viz., inside the convex hull of sensors) in each MC run. The second, on the other hand, completely randomizes the positions of the source and $L=10$ sensors, in the sense that they are all randomly selected from the same square area in MC trials. As summarized in Table \ref{table_MCCTDOA}, our competitors include the semidefinite programming (SDP) based MT method in \cite{ZSu} (termed SDP-TOA), SDP-based robust method for solving Formulation 1 in \cite{GWang3} (termed SDP-Robust-R1), and SDP-based robust method for solving Formulation 2 in \cite{GWang3} (termed SDP-Robust-R2). An overview of their computational complexity is likewise provided in Table \ref{table_MCCTDOA}. The convex programs are handled using the MATLAB CVX package \cite{MGrant}. It is worth noting that the successful application of SDP-Robust-R1 and SDP-Robust-R2 to mitigating the NLOS errors in TDOA SL requires additional \textit{a priori} knowledge about the upper bound on NLOS errors (i.e., uniform distribution parameter $b$ in our case). In contrast, MCC-PNN and SDP-TOA do not premise on such prior information.

\begin{figure*}[!t]
	\centering
	\subfigure[]{\includegraphics[width=2.63in]{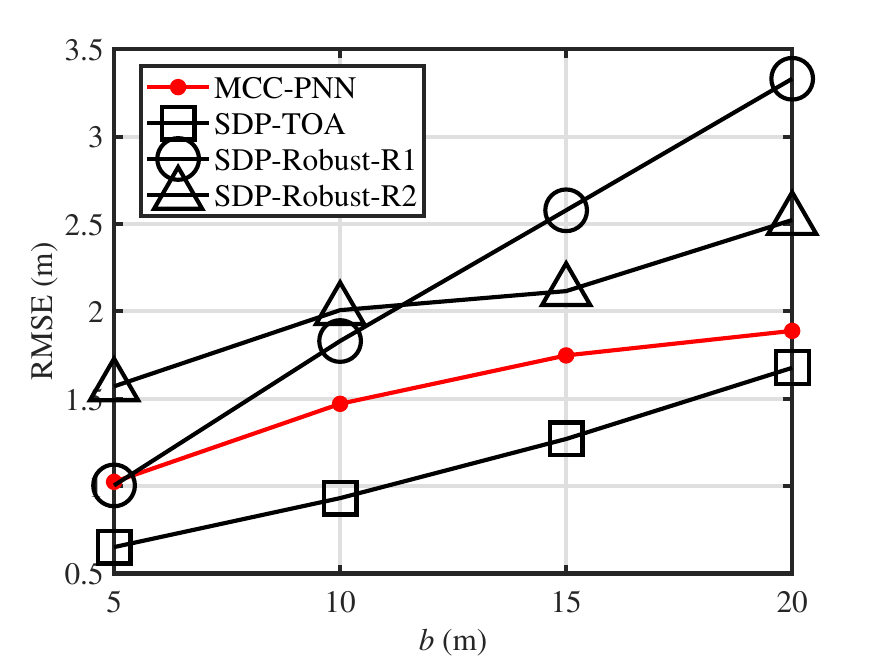}}
	\subfigure[]{\includegraphics[width=2.63in]{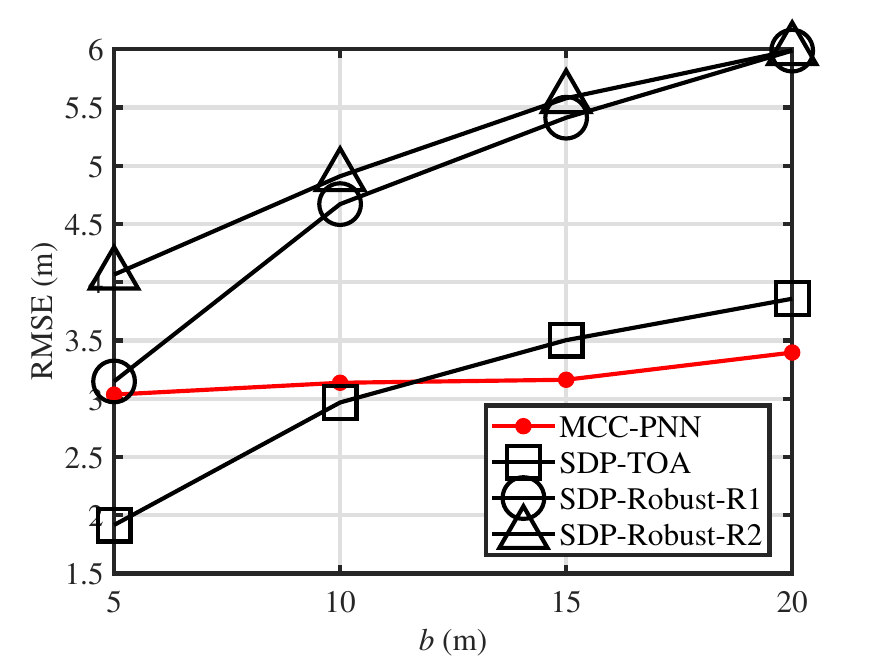}}
	\subfigure[]{\includegraphics[width=2.63in]{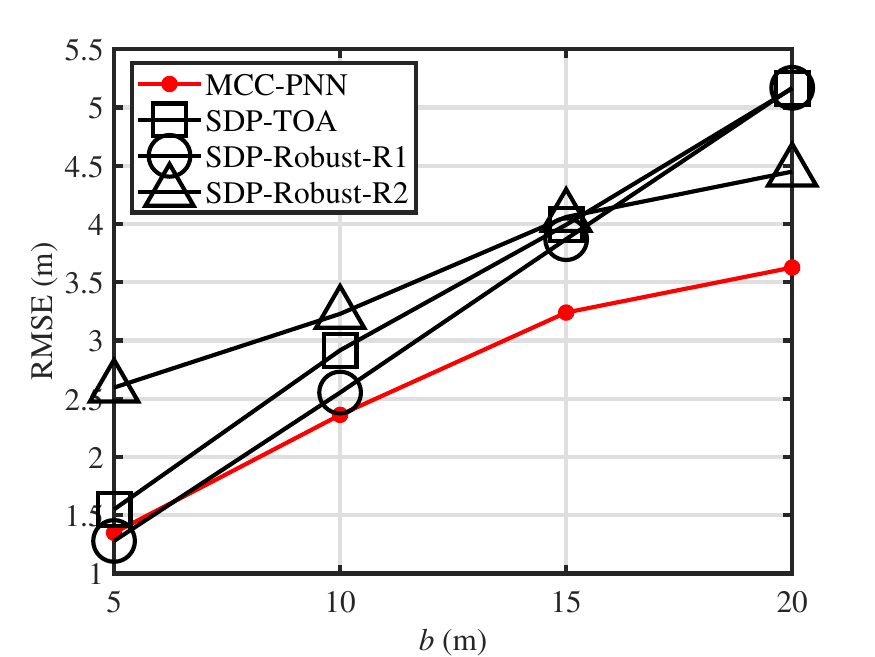}}
	\subfigure[]{\includegraphics[width=2.63in]{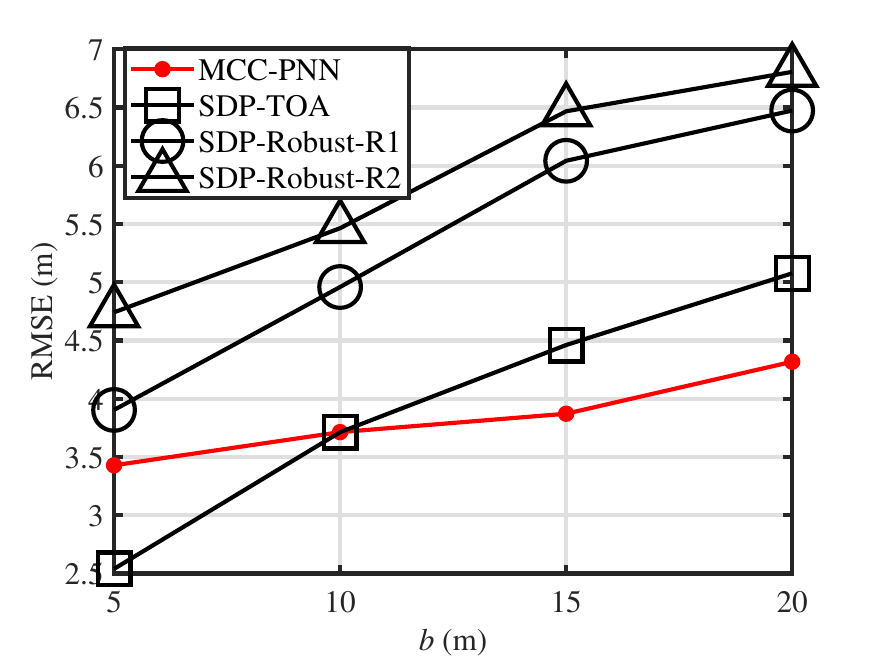}}
	\subfigure[]{\includegraphics[width=2.63in]{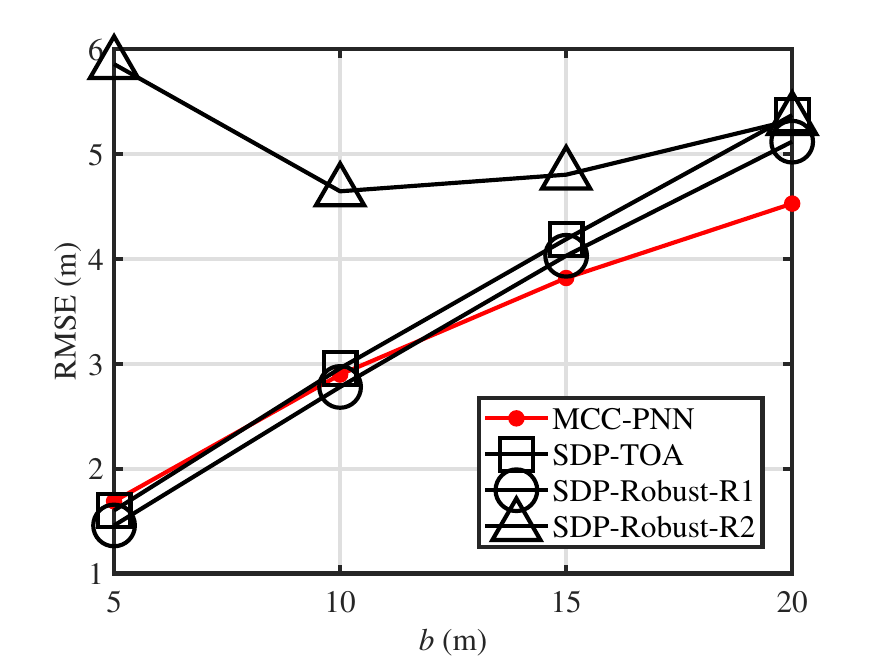}}
	\subfigure[]{\includegraphics[width=2.63in]{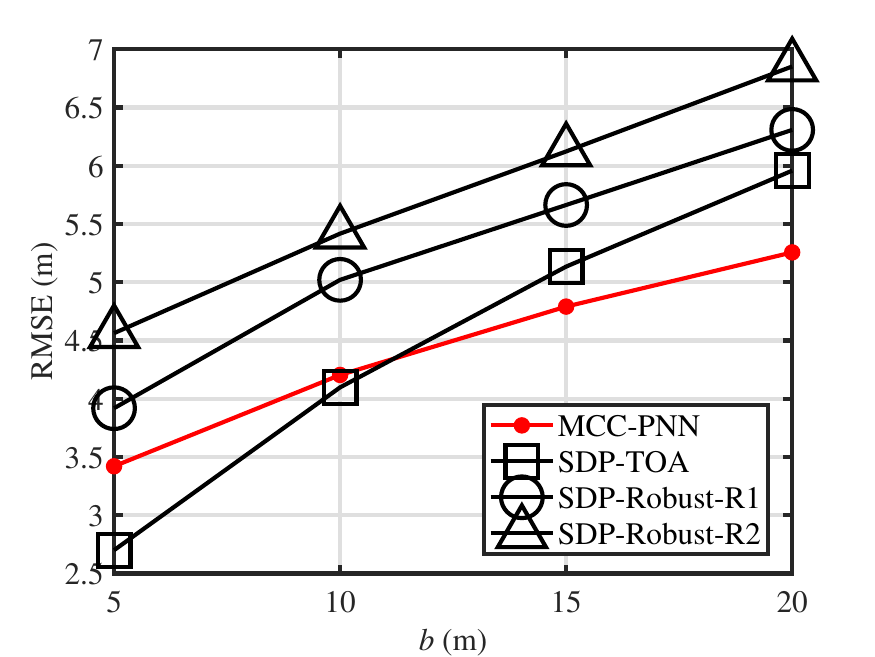}}
	\centering
	\caption{RMSE versus uniform distribution parameter $b$ for two random deployment configurations in mild, moderate, and severe NLOS environments. (a) Randomized source position, $L = 8$, $L_{\textup{NLOS}} = 2$. (b) Randomized source and sensor positions, $L = 10$, $L_{\textup{NLOS}} = 2$. (c) Randomized source position, $L = 8$, $L_{\textup{NLOS}} = 4$. (d) Randomized source and sensor positions, $L = 10$, $L_{\textup{NLOS}} = 5$. (e) Randomized source position, $L = 8$, $L_{\textup{NLOS}} = 6$. (f) Randomized source and sensor positions, $L = 10$, $L_{\textup{NLOS}} = 8$.}
	\label{Figs_RMSE}
\end{figure*}

With the NLOS status in each scenario being specified in the caption, six subplots in Fig. \ref{Figs_RMSE} depict the RMSE versus $b \in [5, 20]$ m for two random deployment configurations again in the representative mild, moderate, and severe NLOS environments. We see that the localization accuracy of all methods (except for SDP-Robust-R2 when $b \in [5, 10]$ m) tends to degrade as $b$ increases. Nonetheless, MCC-PNN generally exhibits the highest level of robustness to the increment of $b$. This is reasonable, and can be explained as the decreasing function of sample correntropy will saturate and become like the $\ell_0$ ``norm'' once the fitting error exceeds a certain threshold (see \cite{WLiu} and Fig. \ref{xiong1_MCCTDOA}). On the contrary, though none of SDP-TOA, SDP-Robust-R1, and SDP-Robust-R2 simply impose the $\ell_2$ loss on the fitting errors $\{ e_{i,1} \}$ or $\{ e_{i} \}$\footnote{Instead of minimizing $\sum_{i=2}^{L} e_{i,1}^2$ or $\sum_{i=1}^{L} e_{i}^2$, they either perform the LS estimation based on the SR model and small Gaussian noise assumption \cite{ZSu} or minimize the supremum of the $\ell_2$ cost function \cite{SBoyd}.}, the corresponding localization formulations still inherit the outlier-sensitivity of the LS criterion to some extent. In terms of the RMSE performance, MCC-PNN is observed to outperform SDP-Robust-R1 and SDP-Robust-R2 in the majority of cases (i.e., except for $b=5$ m in Fig. \ref{Figs_RMSE} (a), $b=5$ m in Fig. \ref{Figs_RMSE} (c), and $b \in [5, 10]$ m in Fig. \ref{Figs_RMSE} (e)). Such results can be rather promising, especially given that MCC-PNN works without additional prior knowledge about NLOS error bounds, which is however required by the robust SDP schemes. In the randomized source location configuration, Figs. \ref{Figs_RMSE} (c) and (e) demonstrate the superiority of MCC-PNN over SDP-TOA in the moderate NLOS environment and most cases in the severe conditions. Figs. \ref{Figs_RMSE} (b), (d), and (f) with the complete randomized source and sensor positions, in comparison, show a boundary of $b=10$ m between scenarios where SDP-TOA delivers lower RMSE than MCC-PNN and the opposite circumstances. The reason why MCC-PNN may perform worse than SDP-TOA when $b$ is small in the magnitude is that SDP-TOA is essentially an estimation-based method, which estimates $q_i$-related variables in addition to the source location, and is known to gain from the comparatively more accurate measurements \cite{HChen2}. On the other hand, our MCC-PNN solution corresponds to a robust method built upon the MCC and can be less sensitive to large NLOS errors than its competitors.

\begin{figure}[!t]
	\centering
	\includegraphics[width=6.5in]{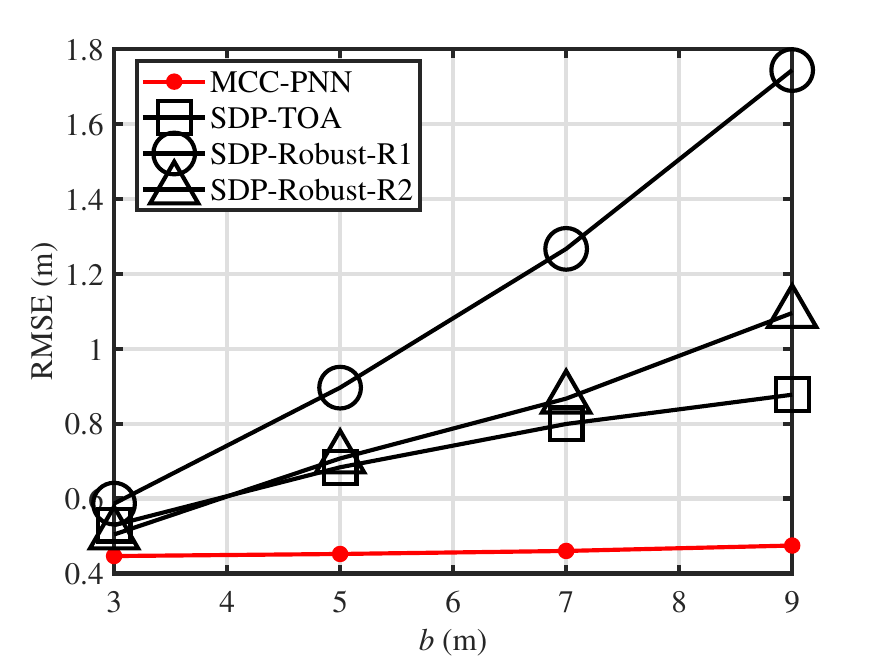}
	\caption{RMSE versus uniform distribution parameter $b$ for deterministic deployment in mild NLOS environment.} 
	\label{Fig_RMSE2}
\end{figure}

Although MCC-PNN is slightly inferior to SDP-TOA over the whole range of $b$ in Fig. \ref{Figs_RMSE} (a), we note that the former can be more beneficial than the latter for certain source positions using the same set of deterministically deployed sensors. For instance, Fig. \ref{Fig_RMSE2} plots the RMSE as a function of $b$ when the true source position and kernel width are fixed as $\bm{x} = [2, 3]^T$ m and $\sigma = 0.8$, respectively (namely, the configuration adopted in Figs. \ref{Figs_DBehaviors} (a) and (b)). It is seen that MCC-PNN achieves the smallest RMSE in such a scenario.

\section{Conclusion}
\label{Sect_CC}
In this paper, we have considered robustifying the traditional TDOA SL formulation via the MCC to mitigate the adverse effects of NLOS propagation. Together with the underlying TOA composition modeled using the newly introduced variable for source onset time and the available sensor-collected timestamps, this leads to a highly nonconvex constrained minimization problem. To perform the optimization, we have proposed a PNN-based neurodynamic approach. We have also verified the local stability conditions of equilibrium for the corresponding dynamical system. The resulting robust TDOA SL scheme does not require any \textit{a priori} NLOS information, and has a quadratic computational complexity in the number of sensors when it is realized discretely. Simulation studies have shown the strong robustness of our method to large NLOS errors and demonstrated its capability of being superior to several recently proposed TDOA techniques in terms positioning accuracy.

\section*{Acknowledgment}
This work was supported by the state graduate funding coordinated by Uni Freiburg.

\end{document}